\documentclass[11pt]{article}
\pdfoutput=1
\usepackage{graphicx,amssymb,amsmath}
\newcommand{\cfig}[2]{\centering\includegraphics[width=#1\textwidth]{figs/#2}}
\newcommand{\fig}[2]{\includegraphics[height=#1\textheight]{figs/#2}}

\begin{document}
\title{\bfseries
The TUS detector of extreme energy cosmic rays on board the Lomonosov satellite
}

\author{%
P.A. Klimov, M.I. Panasyuk, B.A. Khrenov, G.K. Garipov, N.N.~Kalmykov,\\
V.L. Petrov, S.A. Sharakin, A.V. Shirokov, I.V.~Yashin, M.Yu.~Zotov\\
\itshape
Lomonosov Moscow State University, Skobeltsyn Institute of Nuclear
Physics,\\
\itshape
Moscow, Russia\\
S.V. Biktemerova$^a$, A.A. Grinyuk$^a$, V.M. Grebenyuk$^{a,b}$, 
M.V.~Lavrova$^a$,\\ 
L.G. Tkachev$^{a,b}$, A.V. Tkachenko$^a$\\
\itshape
$^a$Joint Institute for Nuclear Research, Dubna, Moscow Region, Russia\\
\itshape
$^b$Dubna State University, Dubna, Moscow region, Russia\\
I.H. Park, J. Lee, S. Jeong\\
\itshape
Department of Physics, Sungkyunkwan University, Seobu-ro, Jangangu,\\
\itshape
Suwonsi, Gyeongido, 440-746, Korea\\
O. Martinez, H. Salazar, E. Ponce\\
\itshape
Benem\'{e}rita Universidad Aut\'{o}noma de Puebla,\\
\itshape
4 sur 104 Centro Hist\'orico C.P. 72000, Puebla, Mexico\\
O.A. Saprykin, A.A. Botvinko, A.N. Senkovsky, A.E. Puchkov\\
\itshape
Space Regatta Consortium, ul. Lenina, 4a, 141070, Korolev,\\
\itshape
Moscow region, Russia
}

\date{%
	Corresponding authors: Pavel Klimov (pavel.klimov@gmail.com),
	I.H.Park~(ilpark@skku.edu)
}
\maketitle

\begin{abstract}
	The origin and nature of extreme energy cosmic rays (EECRs), which have
	energies above the $5\cdot10^{19}$~eV, the Greisen--Zatsepin--Kuzmin (GZK)
	energy limit, is one of the most interesting and complicated problems in
	modern cosmic-ray physics. Existing ground-based detectors have helped
	to obtain remarkable results in studying cosmic rays before and after
	the GZK limit, but have also produced some contradictions in our
	understanding of cosmic ray mass composition. Moreover, each of these
	detectors covers only a part of the celestial sphere, which poses
	problems for studying the arrival directions of EECRs and identifying
	their sources. As a new generation of EECR space detectors, TUS
	(Tracking Ultraviolet Set-up), KLYPVE and JEM-EUSO, are intended to
	study the most energetic cosmic-ray particles, providing larger, uniform
	exposures of the entire celestial sphere. The TUS detector, launched on
	board the Lomonosov satellite on April~28, 2016, from Vostochny
	Cosmodrome in Russia, is the first of these. It employs a single-mirror
	optical system and a photomultiplier tube matrix as a photo-detector and
	will test the fluorescent method of measuring EECRs from space.
	Utilizing the Earth's atmosphere as a huge calorimeter, it is expected
	to detect EECRs with energies above $10^{20}$~eV.  It will also be able to
	register slower atmospheric transient events: atmospheric fluorescence
	in electrical discharges of various types including precipitating
	electrons escaping the magnetosphere and from the radiation of meteors
	passing through the atmosphere. We describe the design of the TUS
	detector and present results of different ground-based tests and
	simulations.
\end{abstract}

%\noindent
%Keywords: extreme energy cosmic rays, transient atmospheric events,
%space fluorescence detectors.

\section{Introduction}

The extremely low flux of extreme energy cosmic rays (EECRs) with
energies above $\sim50~\text{EeV}=50\cdot10^{18}$~eV (the
Greisen--Zatsepin--Kuzmin (GZK) energy limit) prevents the collection of
sufficiently large amounts of data by existing ground-based experimental
arrays. One solution for this problem is the development of new
detection methods with exposures of at least an order of magnitude
greater than those achieved by the existing approaches. The observation
of the ultraviolet (UV) fluorescence of extensive air showers (EAS) from
satellites proposed by Benson and Linsley (1981) promises to become such
an approach. Such a large-aperture, space-based fluorescence detector
situated in an orbit above the atmosphere ($R\sim400$--500~km) should
provide a field of view (FOV) sufficiently wide to observe area of the
atmosphere of $\sim10^5$~km$^2$ in order to collect an adequate sample of
fluorescence photons; in comparison, the largest area of the fluorescent
ground-based detectors is the 3000~km$^2$ Auger project (Abraham et al.,
2010). One technical challenge in the development of an optical system
for such a space detector is the development of sufficient precision
optics ($~1$~km resolution in the atmosphere) and large aperture (2--3~m in
diameter) with an FOV of $\pm15$--30 degrees. Two alternative optical
systems have been suggested for space fluorescence detectors:

\begin{enumerate}
	\item
	Wide-field-of-view optics implemented with complex large-lens optical
	systems. This approach was chosen for the EUSO (Scarsi, 1997) and
	JEM-EUSO (Takahashi et al., 2009) projects.

	\item
	Large-mirror optics-based technology from the development of
	large-area concentrators for solar generators (a Russian initiative of
	SINP MSU and RCS Energia, see Garipov et al., 1998). Such an optical
	system was realized in the TUS detector (Abrashkin et al., 2007) and is
	being implemented with a few improvements for the KLYPVE project
	(Aleksandrov et al., 2000; Khrenov et al., 2001; Panasyuk et al., 2015).
	An optical scheme with a huge mirror has also been considered for the
	OWL project (Stecker et al., 2004; Krizmanic et al., 2013).
\end{enumerate}

Orbital EECR detectors have the following advantages over ground-based
experiments:

\begin{enumerate}
	\item
	Because of their great distance from the detector and the detector's
	high resolution, EECR particle tracks can be observed over huge
	atmospheric areas. Given the height of the Lomonosov satellite orbit
	($R\sim500$~km), TUS will survey areas of up to 6400~km$^2$.

	\item
	After several years of in-orbit operation, a single detector will
	make a uniform observation of the entire celestial sphere. Despite a
	possible inaccuracy in the determination of primary particle energies,
	this will allow a study of the distribution of EECR arrival directions.
	An unavoidable difference in energy response of ground-based arrays
	causes difference in EECR intensity because sky regions are covered
	differently by these arrays.

\end{enumerate}

At the same time, a space EECR detector meets a number of technical
challenges:

\begin{itemize}

	\item Observations of EAS from distances approximately 10 times
		greater than those of the ground-based experiments require higher
		sensitivity and greater angular resolution. A desirable resolution
		of the detector (FOV of a pixel) should be equal to the diameter of lateral
		electron distribution in a shower. For a satellite orbit height of
		500~km, the angular resolution of an orbital detector should be
		0.4--2~mrad, an order of magnitude greater than the 20~mrad
		resolution of existing detectors.

	\item The night atmosphere background in the UV wavelength band
		(300--400~nm) varies over a satellite route. Data obtained with
		the Universitetsky-Tatiana and Universitetsky-Tatiana-2 satellites
		(Tatiana and Tatiana-2 in what follows) (Sadovnichy et al., 2011;
		Vedenkin et al., 2011; Garipov et al., 2013) assessed the scale of
		such variations as
		$3\cdot10^7$--$2\cdot10^8$~photon~cm$^{-2}$~sr$^{-1}$~s$^{-1}$ for
		moonless nights (the lower value above the oceans, the higher
		above cities and regions of increased nighttime airglow such as
		those encountered during equatorial arcs). However, the upper
		limit increases to
		$2\cdot10^9$~photon~cm$^{-2}$~sr$^{-1}$~s$^{-1}$ during full moon
		nights and auroral activity. Ground-based arrays operate on
		moonless nights at specially chosen locations with noise levels
		not exceeding $5\cdot10^7$~photon~cm$^{-2}$~sr$^{-1}$~s$^{-1}$.

	\item Impulsive noise from lightning and accompanying high altitude
		discharges will add to the average noise level and cause higher
		false triggering rates.

	\item The technology of orbital fluorescence detectors should satisfy
		the complex conditions of operation on board satellites.

\end{itemize}

Bearing such circumstances in mind, a program for a gradual conversion
from a ground-based fluorescence detector to space detectors was
initiated, with the launch of the TUS detector as a first, comparatively
simple instrument in order to prove the suitability of both the optical
system design and the photodetector for satellite operation (Khrenov et
al., 2001).

The TUS detector will also measure other phenomena of transient
atmosphere radiation caused by electrical discharges in the atmosphere,
meteorites, and dust grains with high sensitivity and temporal
resolution, see Khrenov and Stulov (2006), Morozenko (2014), Panasyuk et
al. (2016).

\section{Orbital detector TUS}

The TUS detector on board the Lomonosov satellite (Fig.~1) consists of
the following elements: the segmented mirror-concentrator (SMC), the
photodetector (PD), the photodetector moving system (PDMS), the Solar
light sensor (SLS).
Technical parameters of TUS are presented in Table~1.
  
\begin{figure}[!ht]
	\cfig{.6}{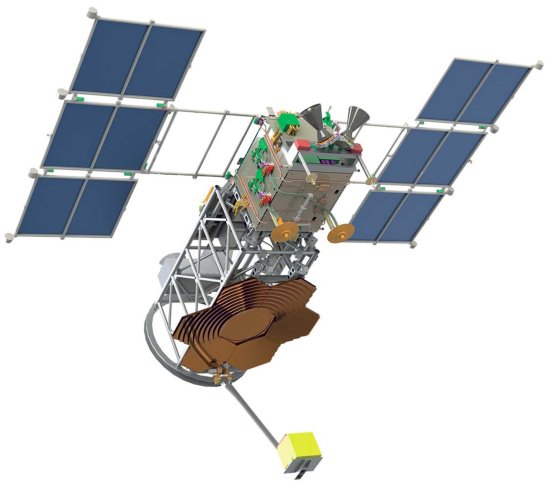}
	\caption{The TUS detector on board the Lomonosov satellite.}
\end{figure}

\begin{table}[!ht]
	\caption{Technical parameters of TUS}
	\begin{center}
	\begin{tabular}{|l|l|}
		\hline
		Parameter & Value \\
		\hline
		Mass & 60 kg \\
		Power (maximum) & 65 W \\
		Data (maximum) &
		250 Mbyte/day \\
		FOV &
		$\pm4.5$ degree \\
		Number of pixels &
		256 (16 clusters of 16 PMTs) \\
		Pixel size (FOV) &
		10 mrad ($5~\text{km}\times5~\text{km}$) \\
		Mirror area &
		2.0 m$^2$ \\
		Focal distance &
		1.5 m \\
		Duty cycle &
		30\% \\

		\hline
	\end{tabular}
	\end{center}
\end{table}

The mirror-concentrator shown in Fig.~2 has an area of 2.0~m$^2$ and is
a Fresnel-type parabolic mirror composed of a central parabolic mirror
and 11 ring-shaped sections. These sections are distributed across 6
hexagonal segments equivalent in size and surrounding the central
segment, which all focus a parallel beam to a single focal point. Thus
the mirror, which has a focal length of 1.5~m, consists of 7 hexagonal
segments, each with a diagonal of 63~cm (Fig.~3, left). These segments
consist of carbon plastic strengthened by a honeycomb aluminium plate
(Fig.~3, right). The mirror construction remains stable over a wide
range of temperatures. In this design, the thickness of the segments is
small (1~cm), which is important for implementation of the mirror on the
satellite frame. The segments were manufactured as plastic replicas of
an aluminium mold (one for the central segment and one for each of the 6
lateral segments).
                                        
% Fig2
\begin{figure}[!ht] 
	%\centerline{\fig{.39}{mirror}\quad\fig{.58}{frame}}
	\centerline{\fig{.3}{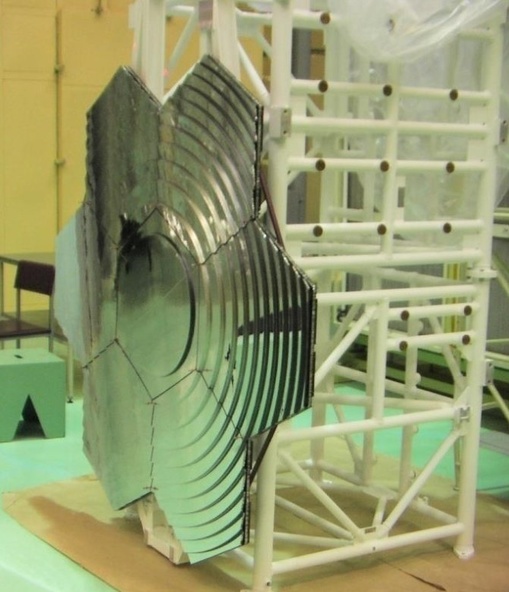}\quad\fig{.3}{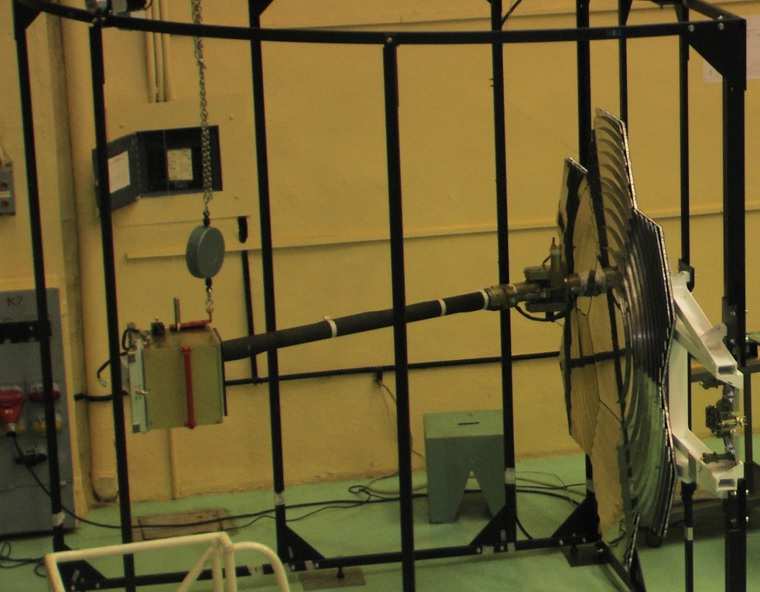}}

	\caption{The TUS segmented mirror-concentrator on
	the Lomonosov scientific payload frame (left) and the TUS detector
	during one of the tests (right)}
\end{figure}

% Fig3
\begin{figure}[!ht] 
	\centerline{\fig{.216}{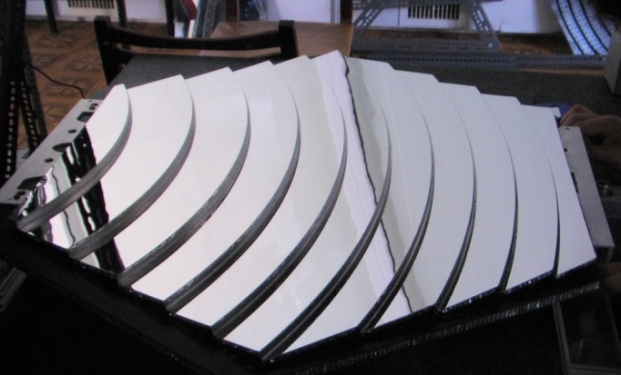}\quad\fig{.216}{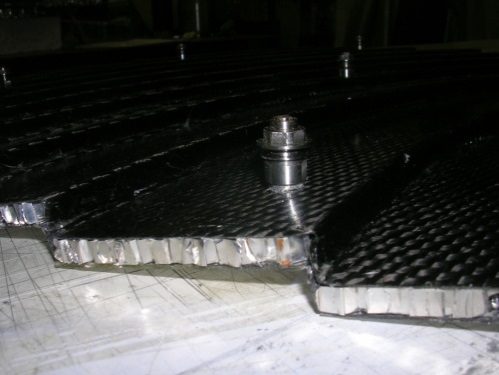}}

	\caption{A lateral segment of the mirror (left) and its three-layer
	structure (right)}
\end{figure}

The plastic mirror surface is covered by an aluminium film and protected
by a MgF$_2$ coat deposited through a vacuum evaporation process. Its
reflectivity at a wavelength of 350~nm (average for atmosphere
fluorescence) is 85\%. The mirror passed various space qualification and
optical tests, which demonstrated the stability of its optical quality
in space conditions. The expected life time of the mirror exceeds~3
years.

The TUS photodetector is a composite of 256 channels (pixels)
positioned at the focal point (Fig.~4). The pixels themselves are
Hamamatsu R1463 photomultiplier tubes (PMTs) with a 13~mm diameter
multialkali cathode. Its quantum efficiency is about 20\% for the 350~nm
wavelength. The multialkali cathode was chosen in place of the bialkali
unit traditionally used in ground-based fluorescence detectors because
of its linear performance over a wider range of temperatures. Special
light guides with square entrance apertures
($15~\text{mm}\times15~\text{mm}$) and circular outputs were employed to
uniformly fill the detector's field of view with PMT pixels. Sixteen
PMTs are combined into an individual cluster (photodetector module), and
each of the 16 clusters of the photodetector has its own digital data
processing system for the first-level trigger, based on a Xilinx FPGA as
well as a high-voltage power supply controlled by the FPGA to adjust the
PMT gain to the intensity of UV radiation. One of the clusters is shown
in the right panel of Fig.~4.

% Fig4
\begin{figure}[!ht] 
	\centerline{\fig{.28}{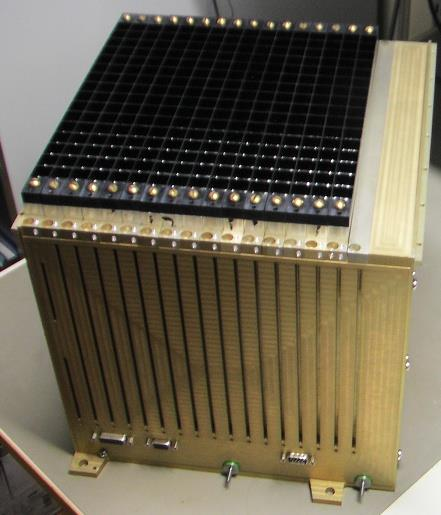}\quad\fig{.28}{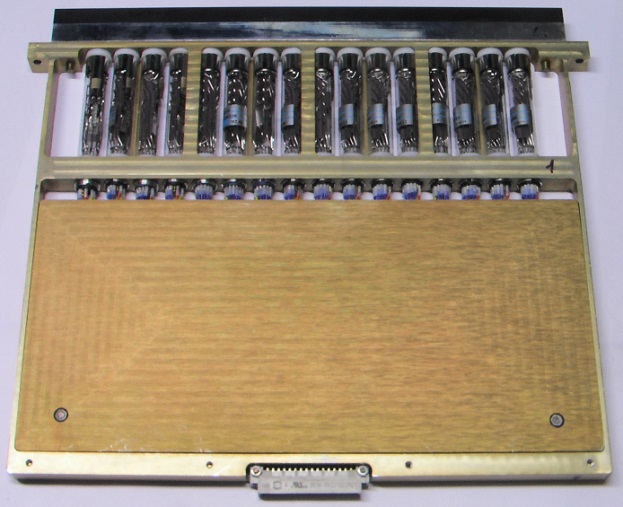}}
	\caption{The TUS photodetector (left) and one of the photodetector
	clusters (right)}
\end{figure}

In parallel to the photodetector measurements, the intensity of light
coming from the Earth is measured by the Solar Light Sensor (SLS), which
is placed on the frame next to the mirror surface and pointed toward the
nadir. The SLS consists of two sensitive photodiodes and electronics
(Fig.~5), and its information is sent to the satellite Information Unit
once per second.
The Information Unit issues a command to
the PDMS to move the photodetector out of the mirror's focus
in case of a dangerous increase in light intensity
resulting from direct sunlight.

% Fig5
\begin{figure}[!ht] 
	\cfig{.3}{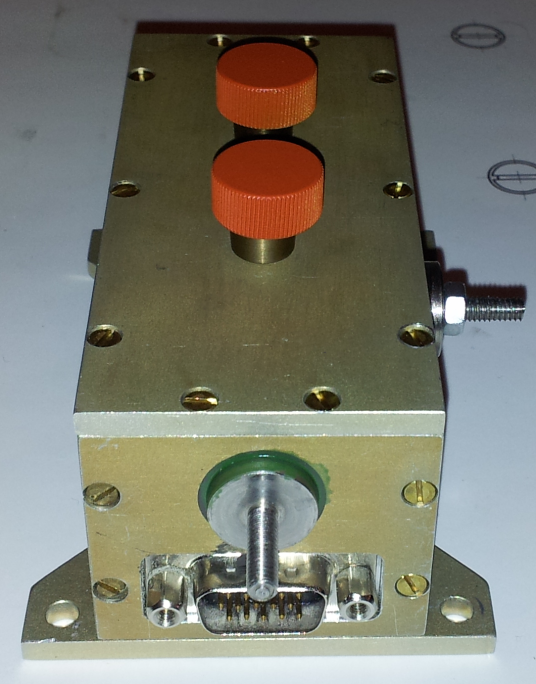}
	\caption{The solar light sensor. The red caps are protective covers
	positioned over the two photodiodes. The ground and Information Unit
	connectors are visible on the front panel of the device}
\end{figure}

\subsection{Digital electronics algorithms}

The TUS detector was designed as a multifunctional orbital set-up that
permits the acquisition of data from various fast (transient) optical
phenomena in the atmosphere. As stated above, the chief focus of these
observations is on EECRs, but TUS is also designed to measure lightning
flashes, transient luminous events (TLEs), and micro-meteors. A
characteristic feature of TLEs is a transient light signal with a
duration that ranges from 1~ms (elves), to dozens of ms (sprites,
gigantic blue jets), to hundreds of ms (blue jets), see Pasko et al.
(2011). The duration of meteor events is in the range of 0.1--1~s. TLEs
and lightning flashes have a significant signal in UV as measured
recently by a number of satellites launched by Lomonosov Moscow State
University, and described in the final section of this paper. 

To measure the varying time frames of the different classes of TLEs, the
photodetector electronics algorithm is implemented as a number of
parallel processes: a system of waveform measurements at different time
scales, a two-level triggering system, and a gain control which adjusts
the sensitivity of the PMT to the UV background. All these processes are
implemented on two sets of boards: 16 photodetector module (PDM) boards,
each with 16 PMTs, and the central processor board (CPB), which gathers
information from all the modules and controls their operation.

The sequence of waveforms is formed by the PDM boards and provides four
types of data (digital oscillograms, DOs) as an output: DO EAS, TLE-1,
TLE-2 and METEOR, which correspond to the duration of three distinct
physical processes in the atmosphere: extensive air showers, transient
luminous events, and micro-meteors respectively. Since duration of TLEs
varies from 1 ms to hundreds of milliseconds, two different waveforms
are provided for their measurement. A fast ADC converts analogue signals
of the PMTs to digital with a time sampling $\tau_0=0.8~\mu$s, which
coincides with the time sampling of the fastest oscillogram (DO EAS). 
For slower waveforms, a system of digital adders is provided (Fig.~6):
\[
	S_k^{(i)} = \sum_{m=0}^{M(i)-1} A_{k+m},\quad i=1,2,3,
\]
where $A_k$ is a digitized signal of each channel. The duration of
adding $M(i)$ has been chosen basing on the characteristic duration of
physical processes apparent in the FOV of a single pixel (see Table~2).
The duration of all oscillograms is set to 256 time samples.

% Fig6
\begin{figure}[!ht] 
	\cfig{.8}{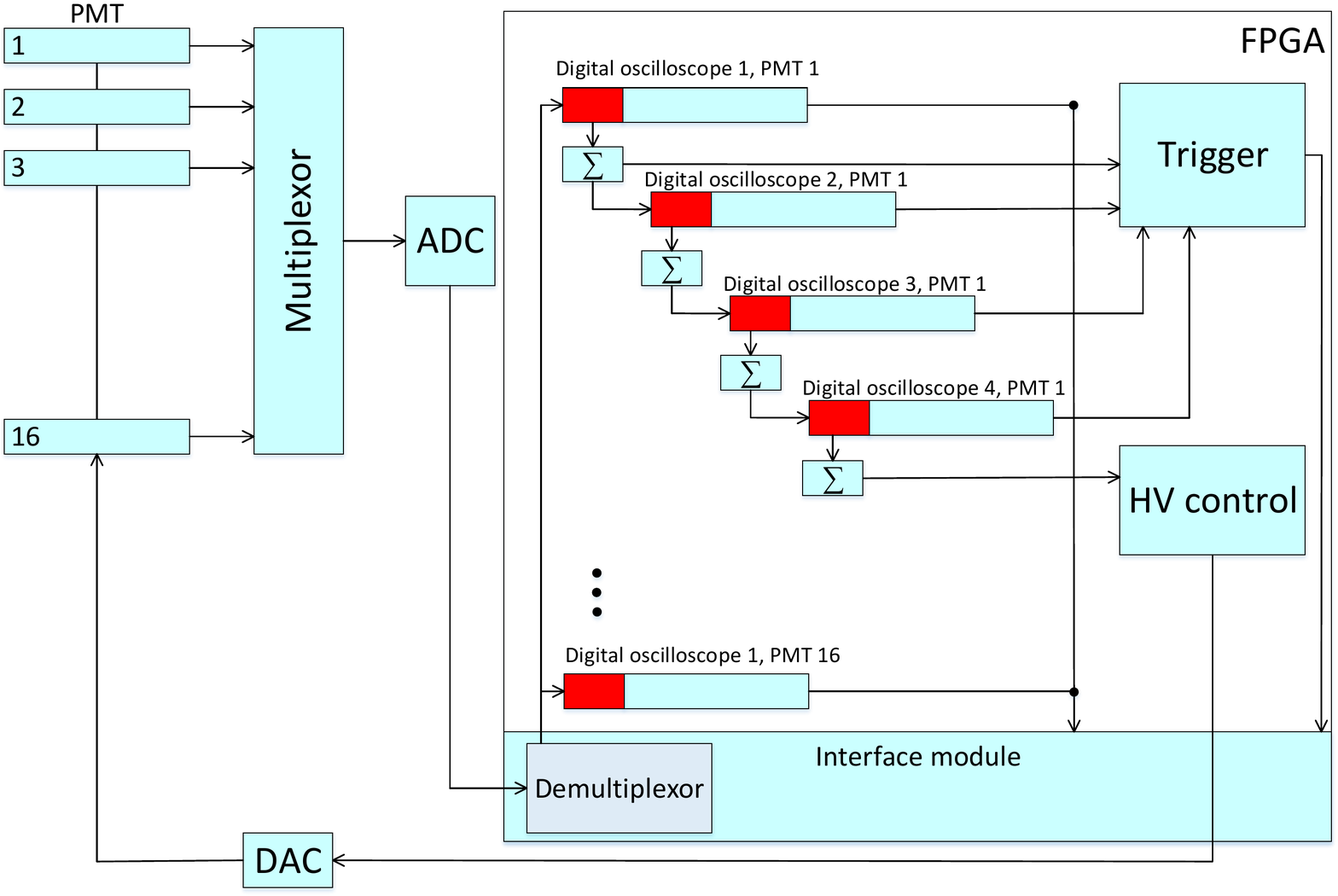}
	\caption{Block scheme of the PDM electronics}
\end{figure}

The first level of the event selection algorithm, a simple threshold
trigger based on a comparison of the signal with a preset level, is
implemented by the individual PDM boards. The signal is integrated
preliminarily over a specified time corresponding to a characteristic
interval set for the development of a phenomenon within a pixel FOV. In
the event of a ``horizontal'' EAS, the time of an EAS movement across a
single pixel is about $15~\mu$s, and an integration time of $t =
2^4\tau_0 = 12.8~\mu$s was chosen. Integration times for other waveforms
are given in Table~2. The preset threshold level for all waveforms can
be changed by a dedicated command from the flight control centre.

\begin{table}[!ht]
	\caption{Temporal characteristics of different DO modes}
	\begin{center}
		\begin{tabular}{|l|c|c|c|}
			\hline
			Digital Oscillogram & DO sampling time & DO length &TS integration time\\
			& $\tau$ & $\Delta T$ & $t$ \\
			\hline
			EAS & $1\tau_0=0.8~\mu$s & $256\tau=205~\mu$s &
			$2^4\tau = 12.8~\mu$s \\
			TLE-1 &
			$2^5\tau_0  = 25.6~\mu$s &
			$256\tau = 6.6$~ms &
			$2^3\tau = 0.2$~ms \\
			TLE-2 &
			$2^9\tau_0  = 0.4$~ms &
			$256\tau = 105$~ms &
			$1\tau = 0.4$~ms \\
			METEOR &
			$2^{13}\tau_0 = 6.6$~ms &
			$256\tau = 1.7$~s &
			$2^4\tau = 105$~ms\\
			\hline
		\end{tabular}
	\end{center}
\end{table}

To decrease the false trigger rate resulting from background
fluctuations, a second trigger level was developed: a pixel mapping
trigger implemented in the CPB, which acts as a contiguity trigger. This
procedure selects cases of sequential triggering of spatially contiguous
pixels (channels) that are also adjacent in time, allowing for the
selection of events with different spatial-temporal patterns. An
additional parameter important for this trigger is the so-called
adjacency length, i.e., the number of neighbouring channels ($N$)
sequentially activated by a signal from a given event. The preset value
of $N = 3$ can be changed by a special command during the flight. All data
(four types of waveforms) are permanently stored in the BRAM of every
PDM FPGA. The PDMs wait for a trigger command from the TUS CPB FPGA
before sending this information to the memory of the CPB, which, in
turn, relays the triggered oscillograms (the data of all 256 pixels over
256 time intervals) along with high voltage and time data to the
Information Unit via the CAN bus interface. One frame of data (one
event) is expected to be about 100 Kbytes, and the limit of TUS data
determined by a dedicated Information Unit memory is about 250
Mbytes/day.

UV radiation measurements are performed in the DC mode, and the signal
is integrated in the anode RC-chain of each PMT, with the time constant
set to 600~ns. Measurements of UV intensity allow adjusting the PMT gain
control via the high voltage (HV) control system to avoid saturation of
the PMTs under conditions of increased UV intensity, such as during
moonlit nights or transits across regions of greater UV emissions
(auroral regions, large cities, etc.). Two algorithms are implemented in
the PDM FPGA to tune the HV and thus solve the problem: (i) the base
level of the ADC code is kept constant, or (ii) the first level trigger
rate is calculated and kept constant. The second algorithm allows a
larger dynamic range, while the first is more reliable in the event of
errors or high trigger rates caused by defective PMTs. The HV is
controlled by the output voltage level of a dedicated DAC (see Fig.~6.),
and HV correction occurs once every 100 ms to ensure a constant voltage
during EAS and TLE oscillogram output. During a METEOR oscillogram, the
HV is adjustable up to 16 times.

\subsection{Photodetector tests}

The first qualification test of a Hamamatsu R1463 PMT was performed with
the same hardware and software complex (test bench) that had been
utilized successfully on a previous occasion by the Joint Institute for
Nuclear Research (JINR, Dubna) in testing PMTs for the LHC's ATLAS Tile
Calorimeter. After the test, PMTs with nearly equal gain were grouped
into 16 tube clusters (modules). It is important to note that the 16
PMTs of each separate module have a common HV power supply, and their
gain was equalized by adjusting the resistors of the individual tubes'
voltage dividers. PMTs within a single module were adjusted to have
identical gain across the entire range of the HV control for local night
times (DAC codes 160--250). Tests with a reference light source were
performed after adjusting the PMTs across the entire range of HV
variation. The results of these tests on one of the PMT clusters are
presented in Fig. 7. Note that the characteristics of all the PMTs
within this cluster drew closer to each other as a result of their
adjustment.

% Fig7
\begin{figure}[!ht] 
	\cfig{.75}{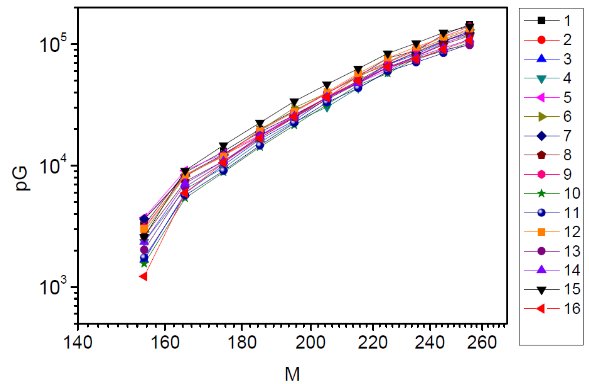}

	\caption{Results of adjusting PMTs in one of the PMT modules. Here,
	$p$~is the photocathode quantum efﬁciency, $G$~is the PMT gain, and $M$is
	the DAC code}
\end{figure}

During space operation, each of the 256 PMTs measures the background
intensity of atmospheric UV. For an average background, distribution of
pixel signals in the photodetector is determined by the distribution of
pixel gains. A map of 256 pixel signals related to the standard signal
from a UV source was measured before launch and presented in Fig.~8.
Note that there is still a large difference in the sensitivity of
specific PMTs in different modules. The obtained deviation from the
reference signal was not satisfactory for selecting an EAS because a few
pixels with a high gain could produce a major portion of selected
events. To reduce the width of the gain distribution, signals from the
reference source were corrected digitally and then sent to the trigger
system.

% Fig8
\begin{figure}[!ht]
	\cfig{.75}{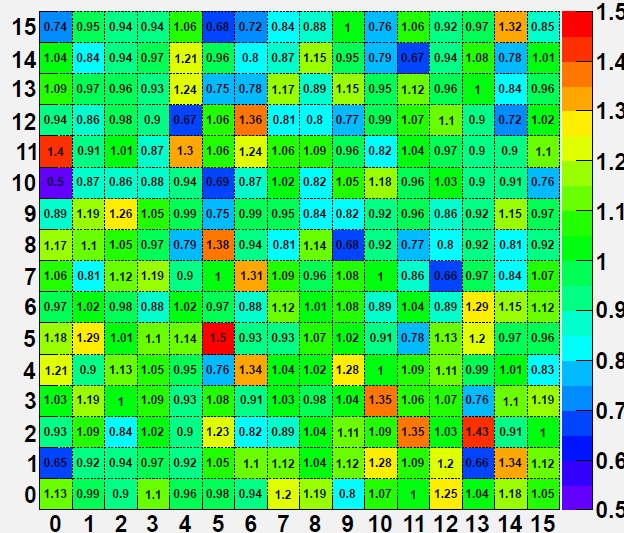}
	\caption{Map of the TUS pixel gains after preliminary PMT grouping
	and adjustment. Numbers in the boxes are PMT signal values for the
	reference standard UV pulse. Refer to the text for details}
\end{figure}

The TUS photodetector successfully passed autonomous thermal vacuum
tests at Skobeltsyn Institute of Nuclear Physics of Lomonosov Moscow
State University as well as a series of complex tests in a vacuum
chamber at VNIIEM Corporation (JSC).

\subsection{TUS performance}

Performance of the TUS detector was simulated by Grinyuk et al. (2013)
with the ESAF software framework (Fenu et al., 2011), taking into
account parameters of the real TUS mirror-concentrator and TUS
electronics.

The focusing of the mirror-concentrator was checked through experimental
measurements of the mirror point spread function (PSF). However
experimental PSF determinations differ from the PSF of an ``ideal''
mirror, which is free of the technological defects that appear during
the production of the actual mirror. In Fig.~9, results of the real
mirror PSF measurement (right panel) are compared with the PSF of the
``ideal'' mirror (left panel). In the real measurements, the light beam
was tested at 8 different azimuthal and four polar angles
$\theta=0^\circ$, $1.5^\circ$, $3^\circ$, $4.5^\circ$. As indicated in
Fig.~9, the real mirror PSF differs from the ideal point value even at
small polar angles. Nevertheless, for a $9^\circ$ diameter field of
view, the PSF is compatible with the TUS pixel size.
                                       
\begin{figure}[!ht]
	\cfig{.99}{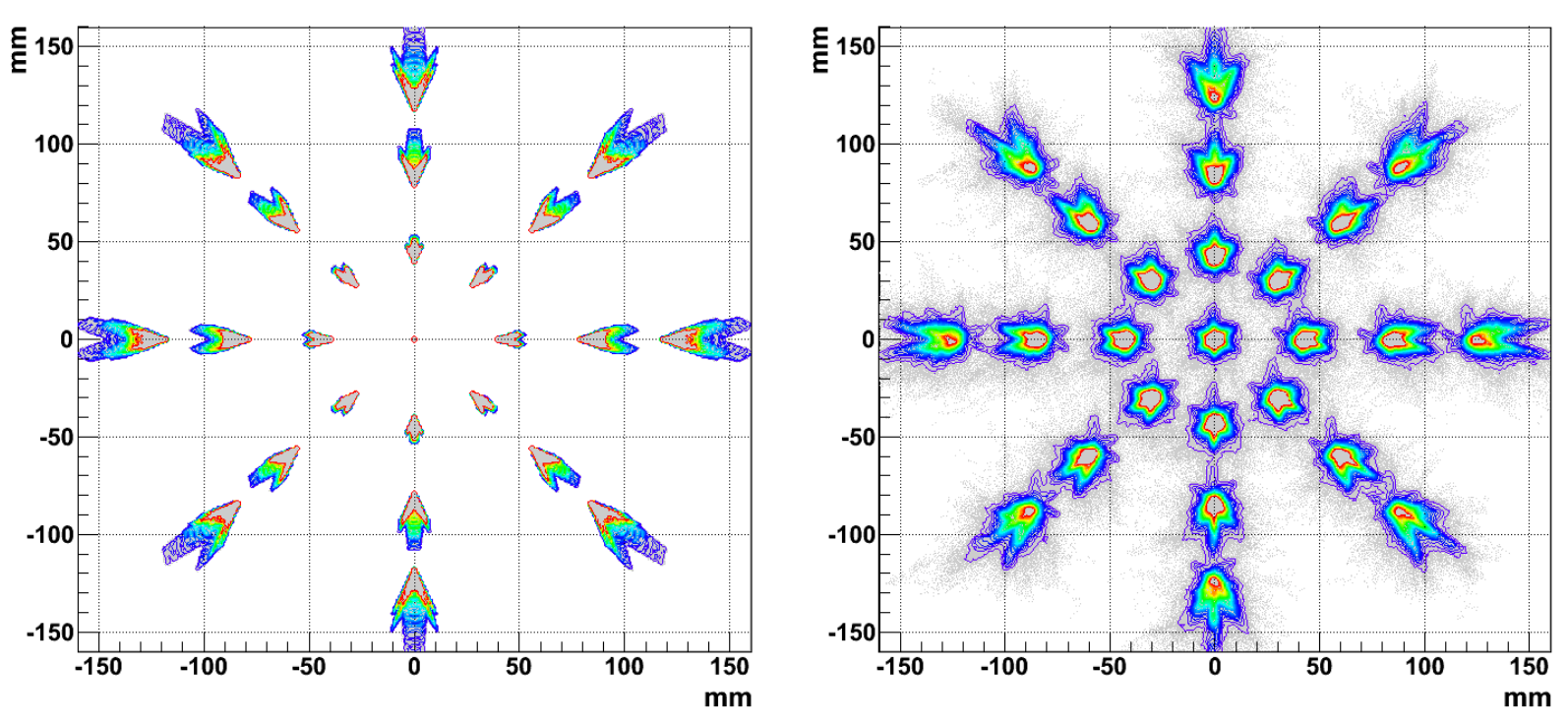}
	%\centerline{\fig{.3}{psf1}\quad\fig{.3}{psf2}}
	\caption{PSF for different azimuthal and polar angles. Left panel: an
	``ideal'' mirror, right panel: the real mirror}
\end{figure}

EAS signals in photodetector pixels (fluorescence photon numbers) were
calculated for the real and ideal mirrors. Typical signals are presented
in Fig.~10, which shows the percentage of photons received by the
mirrors and distributed over pixels for an EAS with a zenith angle of
$75^\circ$. The photon distributions along the EAS cascade curve for the ideal
and real mirrors show little difference, which confirms the high quality
of the TUS mirror.
                                        
\begin{figure}[!ht]
	\centerline{\fig{.24}{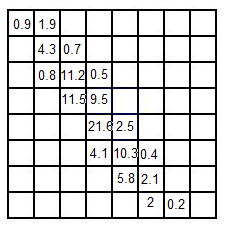}\quad\fig{.24}{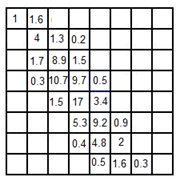}}

	\caption{EAS signals in the photodetector presented as percentage of
	fluorescence photons read by a single pixel to the number of photons
	collected by a mirror of 2~m$^2$ area. Left panel: the ``ideal'' mirror;
	right panel: the real mirror. Only pixels with percentage $>0.2$\% are
	shown}
\end{figure}

To estimate the efficiency of the TUS trigger system and the accuracy of
measuring EAS parameters, EAS pixel signals must be compared with the
level of background noise in the same pixels. Data on the UV background
of the atmosphere were obtained in measurements on board the Tatiana and
Vernov satellites (Sadovnichy et al., 2011, Vedenkin et al., 2011,
Garipov et al., 2013).  The average background level of UV noise over
the night side of the Earth on moonless nights was found to be 
$\nu_0 = 10^8$~photon~cm$^{-2}$~sr$^{-1}$~s$^{-1}$.
The lowest background signal
on moonless nights is around $0.3\nu_0$ (above the South Pacific,
deserts such as the Sahara, and a part of Siberia). The highest level
determined for full moon reached $10\nu_0$.  The expected value of the
pixel noise level was calculated from these experimental data, and the
background photon counts for the three values of background
intensities~$\nu_0$ are presented in Table~3. In measurements of EAS
pixel signals over the time interval~$T$, the standard
deviation~$\sigma$ (square root of $I_N$ values multiplied by~$T$) is
utilized to simulate noise.

\begin{table}[!ht]
	\caption{Level of the background signal in an individual pixel
	(number of photons per $\mu$s) as a function of the UV background
	intensity}
	\begin{center}
		\begin{tabular}{|l|l|l|l|}
			\hline
			$\nu_0$,~photon~cm$^{-2}$~sr$^{-1}$~s$^{-1}$	&
			$3\cdot10^7$ &
			$10^8$ &
			$10^9$ \\
			\hline
			$I_N$, photons $\mu$s$^{-1}$ &
			36 &
			120 &
			1200 \\
			\hline
		\end{tabular}
	\end{center}
\end{table}

Numerical calculations of EAS signals in pixels were compared with the
pixel noise~$\sigma$ to simulate extensive air showers of various
primary energies and zenith angles. The following TUS parameters were
used: a 2.0~m$^2$ mirror area; a solid angle for a single pixel equal to
$10^{-4}$~sr; and a detector system optical efficiency equal to 0.6
(including a mirror reflection coefficient of 0.85 and an efficiency of
collecting photons from the whole mirror to one pixel). The ratio of
various EAS signals to noise~$\sigma$ for exposures of $T=12.8~\mu$s duration are
presented in Table~4.

\begin{table}[!ht]

	\caption{Signal-to-noise ratio for EAS of various primary energies
	and zenith angles. The UV background intensity is taken to be 
	$\nu_0 = 10^8$~photon~cm$^{-2}$~sr$^{-1}$~s$^{-1}$.}
	\begin{center}
		\begin{tabular}{|l|l|l|l|l|l|}
			\hline
			Energy, EeV &
			$\theta=60^\circ$ &
			$\theta=65^\circ$ &
			$\theta=70^\circ$ &
			$\theta=75^\circ$ &
			$\theta=80^\circ$ \\
			\hline
			100 &
			2.2 &
			2.4 &
			2.7 &
			3.0 &
			3.4 \\
			150 &
			3.3 &
			3.6 &
			4.0 &
			4.5 &
			5.1 \\
			200 &
			4.4 &
			4.8 &
			5.4 &
			6.0 &
			6.8 \\
			300 &
			6.6 &
			7.3 &
			8.1 &
			9.1 &
			10.2 \\
			\hline
		\end{tabular}
	\end{center}
\end{table}

Results of numerical calculations indicated that the trigger system
described above will collect EAS with signals above 3--4 sigmas of
noise, which corresponds to primary energies of about 100--150~EeV for
air showers with zenith angles $60^\circ$--$90^\circ$ registered in the
central part of the detector FOV. For events at the edge of FOV, the
efficiency of the trigger system is lower, and the energy threshold
increases to around 150--200~EeV.

\section{Present data on background effects in measuring EECRs from
space}

As stated above, a direct background effect encountered during nighttime
measurements of fluorescent EAS tracks from space is a luminescence of
the nocturnal atmosphere. Experimental results of measurements of this
atmospheric luminescence or ``airglow'' were obtained by the Tatiana-2
satellite (Vedenkin et al., 2011). The data obtained on moonless nights
during the winter of 2009--2010 over the Earth's nightside for latitudes
between $30^\circ$S and $60^\circ$N are given in Fig.~11. The
intensity~$J$
of the atmospheric glow varies over a wide range of 
$J=3\cdot10^7$--$2\cdot10^8$~photon~cm$^{-2}$~sr$^{-1}$~s$^{-1}$.
It is well known that this
atmospheric glow originates in a comparatively narrow layer of the upper
atmosphere (lower ionosphere) at heights of 80--100~km. An orbital
fluorescence detector directed to the nadir detects the atmospheric glow
practically without absorption by the higher layers of the atmosphere.
Ground-based EAS detectors are not able to detect the glow originating
at altitudes in the 80--100~km range because of its strong absorption by
the atmosphere at lower altitudes. For an orbital observation, the
detection of these upper atmospheric emissions increases up to
$\sim10^8$~photon~cm$^{-2}$~sr$^{-1}$~s$^{-1}$
in some places.

%Fig 11
\begin{figure}[!ht]
	\cfig{.9}{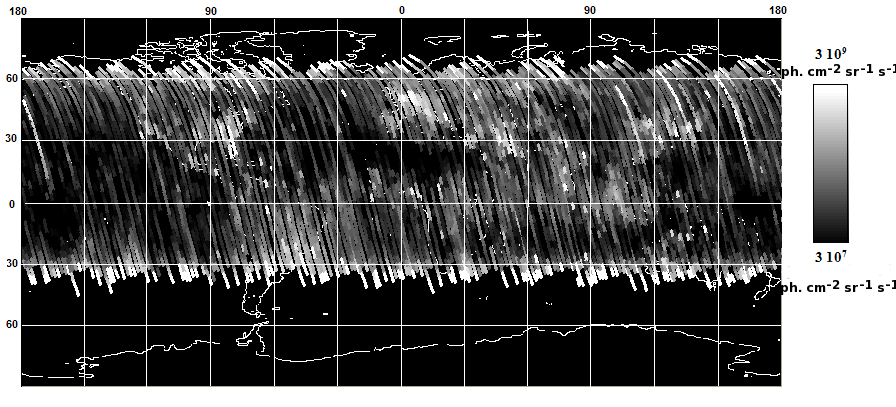}

	\caption{Global map of the nighttime atmospheric glow intensity in
	the UV wavelength band (240--400~nm) as measured by the Tatiana-2
	satellite (Vedenkin et al., 2011)}
\end{figure}

Taking into account the data on the atmospheric glow, the exposure of
TUS with an FOV of $9^\circ$ was estimated by Klimov (2009). The
efficiency of an EECR event selection is close to 100\% for energies $E
> 300$~EeV and those events will be collected with the total exposure of
12000~km$^2$~sr~yr throughout 3 years of in-orbit operation. Events with
energies $E = 70$--300~EeV will be detected with less efficiency
(exposure). This means that for the steep energy spectrum of EECRs above
the GZK limit (the integral spectrum exponent $\sim4$ for energies $E >
50$~EeV), events with energies 70--300~EeV will be selected and measured over
the darkest regions of the Earth: above the Pacific ocean, deserts, and
a part of Siberia (Fig.~12). With such limited exposure, the TUS
detector will not be able to make a breakthrough with regard to the
problem of EECR origins. Nevertheless, the principal aim of the TUS
mission is to test the performance of an orbital EAS fluorescence
detector in a space environment.

%Fig 12
\begin{figure}[!ht]
	\cfig{.5}{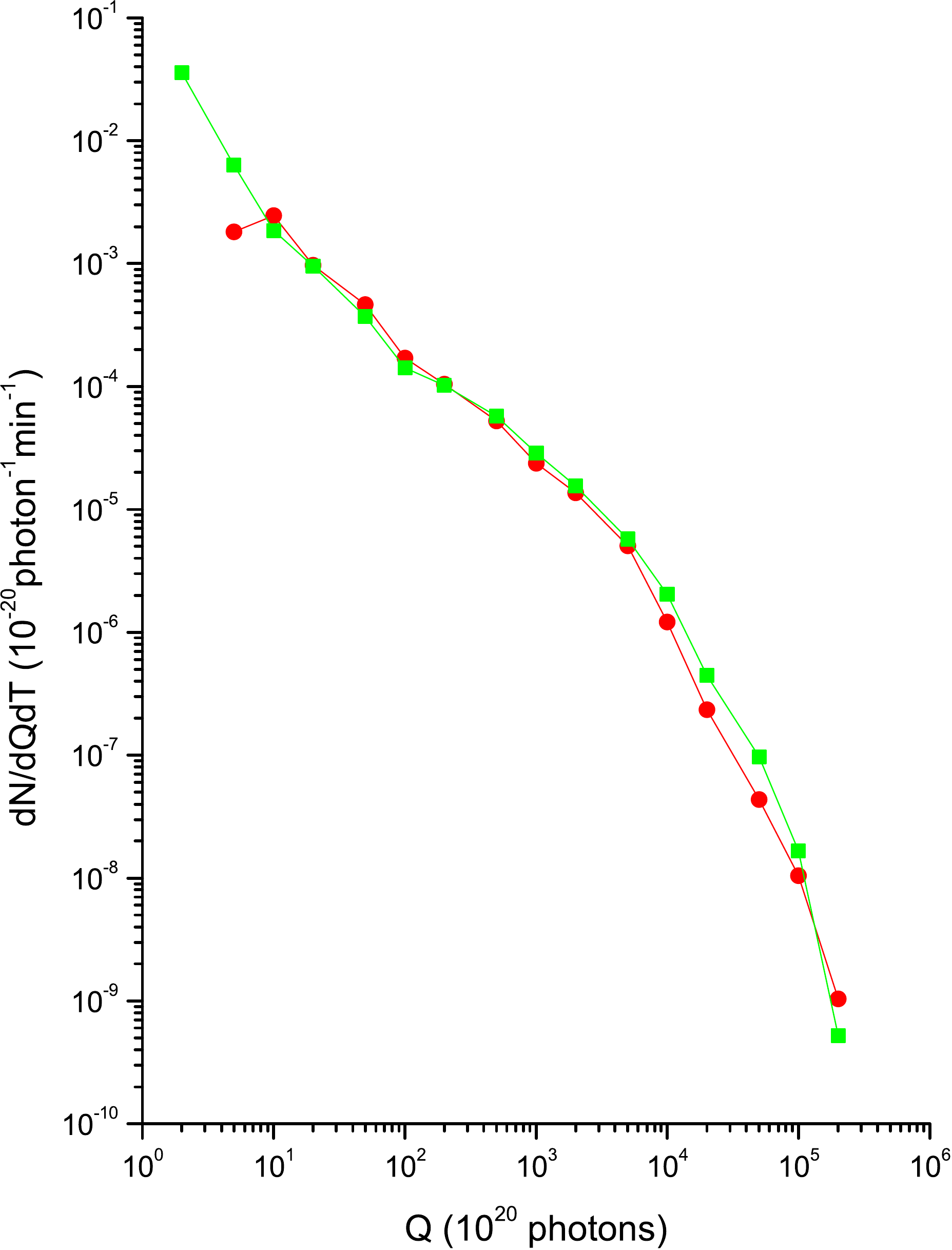}

	\caption{Photon number distribution of UV flashes. Red circles:
	results of the Tatiana-2 experiment (Garipov et al., 2013), green
	squares: the Vernov experiment (Panasyuk et al., 2016)}
\end{figure}

The TUS detector will also operate above intensive sources of background
glow during the flight: auroral events, city lights and other sporadic
lights of uncertain origin. Experimental results of the Tatiana and
Tatiana-2 satellites have demonstrated that these higher intensity glow
sources are limited in area and do not strongly affect detector
exposure.  Another source of background glow in orbital EECR
measurements are short UV flashes (durations of 1--100~ms), the origin of
which is related to electrical discharges in the atmosphere. The latest
results regarding their intensity and distribution were obtained with
the Tatiana-2 satellite (Garipov et al., 2013) and the Vernov experiment
(Panasyuk et al., 2016). The UV detector of these satellites operated in
conditions close to those of an orbital EECR detector, measuring the
temporal structure of flashes over atmospheric regions of thousands
km$^2$, while oriented toward the nadir. Measurements were performed for
a wide range of photon counts~$Q$ per atmospheric UV flash event: from~$Q
=10^{21}$ up to $Q\sim10^{25}$, where tens of events were registered. The main
features of flashes with $Q >10^{23}$ are their duration of 10--100~ms and
the concentration of their global distribution in the equatorial region
over the continents. This suggests that such phenomena are either
lightning flashes or transient luminous events generically related to
lightning. These "bright" flashes will be easily distinguished from EAS
fluorescent signals due to their long duration and enormous photon
counts (to compare, EAS events have durations of less than 0.1~ms and UV
photon counts of $Q\sim10^{16}$ for $E=100$~EeV).

More likely to resemble EECR events are the dim, short flashes ($Q\sim
10^{21}$--$10^{23}$, duration $\sim1$~ms) observed by the Tatiana-2
(Garipov et al., 2013) and Vernov (Panasyuk et al., 2016) satellites.
The flash event distribution over the number of photons~$Q$ is presented
in Fig.~12.  One can see that dim events with small photon counts ($Q <
10^{22}$) constitute a considerable portion of all events, and that the
global distributions of dim and bright flashes measured by Garipov et
al. (2013) were found to differ, see Fig.~13.

% Fig13
\begin{figure}[!ht]                                       
	\cfig{.6}{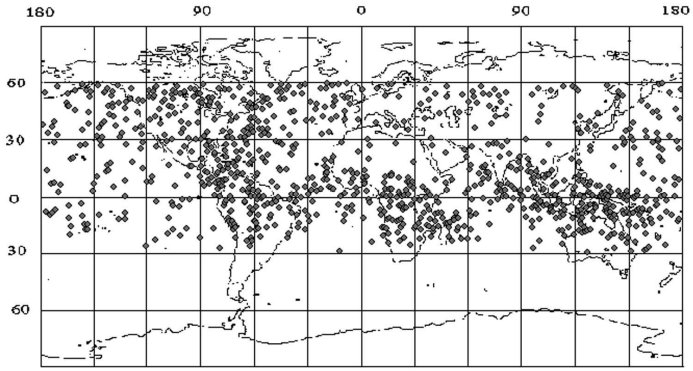}
	\cfig{.6}{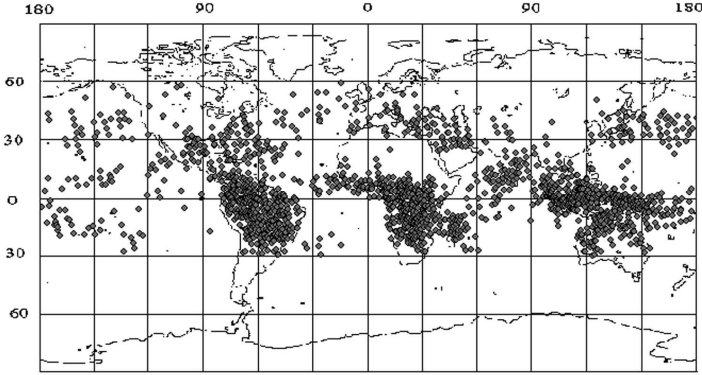}

	\caption{Global distribution of dim (upper panel, $Q < 5\cdot10^{21}$) and
	bright (bottom panel, $Q > 10^{23}$) UV flashes (Garipov et al.,
	2013)}
\end{figure}
  
Bright flashes are concentrated in the equatorial region above
continents, which is consistent with an origination in lightning
activity. Their rate in this region is of the order of 
$10^{-3}$~km$^{-2}$~hr$^{-1}$. 
Dim flashes are distributed more uniformly with a rate of 
$\sim10^{-4}$~km$^{-2}$~hr$^{-1}$. 
The rates of both types of flashes are much higher
than the expected EECR events rates of
$\sim10^{-6}$~km$^{-2}$~hr$^{-1}$. 

The above numbers show the similarity of EECR events to slower
atmospheric flashes, especially during the onset of these slower events.
The TUS mission will provide important data for a better understanding
of these atmospheric background effects.

\section{Data on Transient Atmosphere Events expected from the TUS
detector}

The TUS detector has a number of advantages over previous detectors of
transient atmosphere events (TAEs) on board the Tatiana, Tatiana-2 and
Vernov satellites:

\begin{itemize}

	\item a large aperture (area of the mirror-concentrator $\sim2$~m$^2$) for
		detecting TAE fluorescence in the UV band;

	\item the capacity to measure UV images of TAEs in the 256 pixel
		photodetector with a resolution of 5~km within a field of view of
		80~km$\times$80~km in the atmosphere;

	\item the capacity to measure variation of images in time with the
		digital oscilloscope at four different time scales (see Table~2);

	\item the capacity to select TAEs on four independent time scales.

\end{itemize}

With these advantages, the TUS detector will be able to obtain new data
on TAEs.

\begin{enumerate}

	\item The first result expected from the larger aperture of TUS will
		be an increase in the detection rate of TAEs with low photon
		counts (down to $10^{17}$), which is three orders of magnitude below
		the threshold value for the Vernov experiment (Fig.~12).

	\item The distinction of different TAE types will be improved
		considerably by the imaging and timing of events over expanded
		space and time scales.

	\item It will be interesting to evaluate the possibility of EASs as
		initiators of TAEs, as conjectured by Gurevich and Zybin (2001),
		through observation of the early stages of TAEs. An EAS developing
		in tens of microseconds might initiate a subsequent TAE with a
		duration of tens of milliseconds.

	\item The TUS will detect events repeating on scales ranging from
		milliseconds up to seconds, filling the gap in the data of the
		previous satellite experiments.

	\item Images of TAEs will be also useful in distinguishing upper
		atmosphere events from precipitating electron events initiated by
		a lightning flashes (see Voss et al., 1998). An image of a
		precipitating electron event is expected to be wider than images
		of sprites and will lack the ring shape characteristic of elves.

\end{enumerate}

\section{Results of first measurements}

The EAS mode of operation was mostly employed during the first months of
work in orbit. More than 20 thousand of various events at night parts of
orbits were measured from May till November 2016. They differ in spatial
dynamics and temporal structure of waveforms.

The first registered phenomena that were unexpected are instant and as a
rule intensive flashes that produce linear tracks in the focal surface.
An example is shown in Fig.~14. One can see a flash that occurs during
one time frame simultaneously in a group of PMTs lined up in a track.
Preliminary simulations performed with the Geant4 framework revealed
that such events can be caused by a charged particle penetrating through
the glass filter in front of PMT and producing a significant amount of
fluorescent and Cherenkov light (Klimov et al., 2017).

%14
\begin{figure}[!ht]
	\cfig{.6}{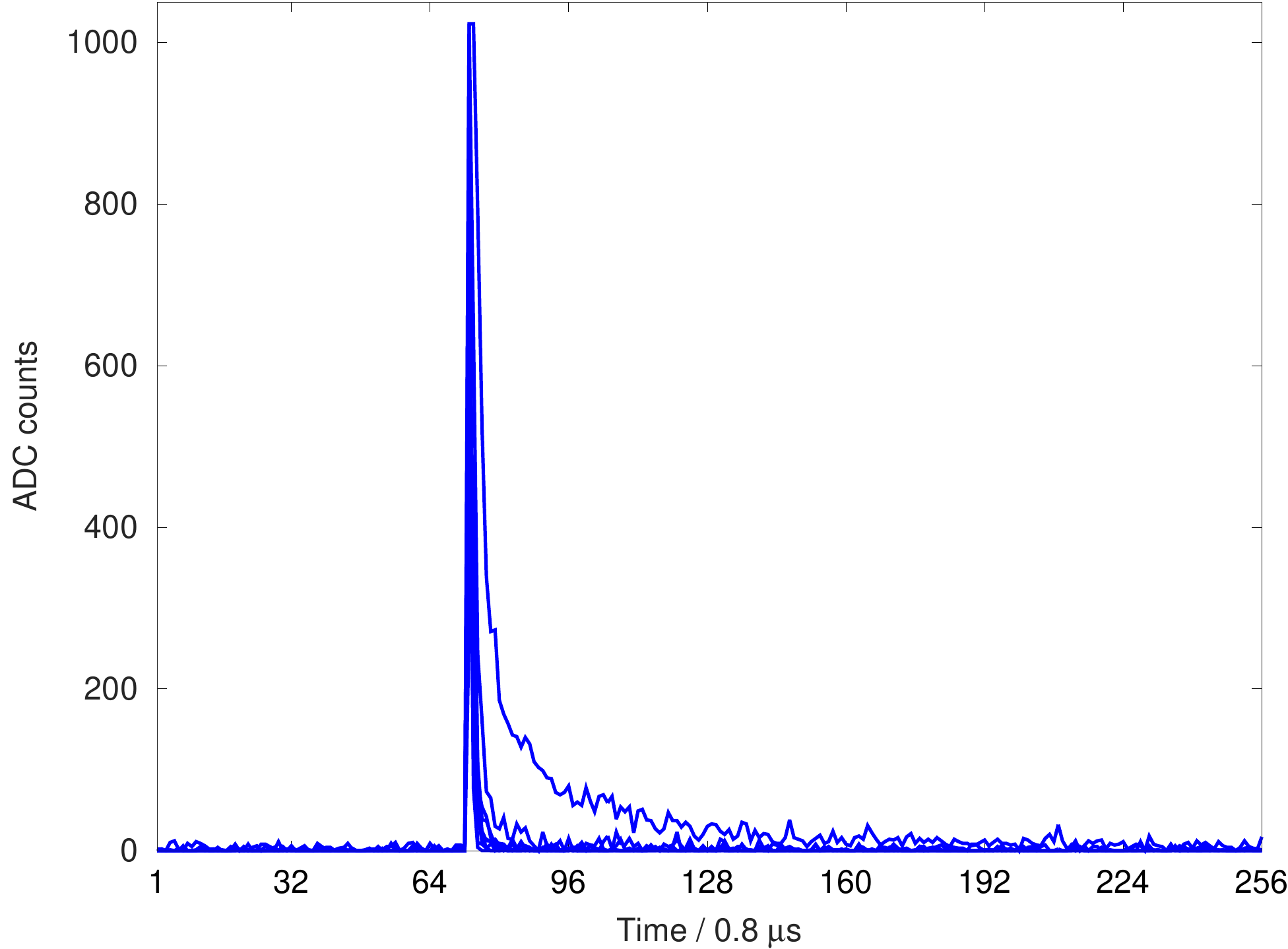}
	\cfig{.5}{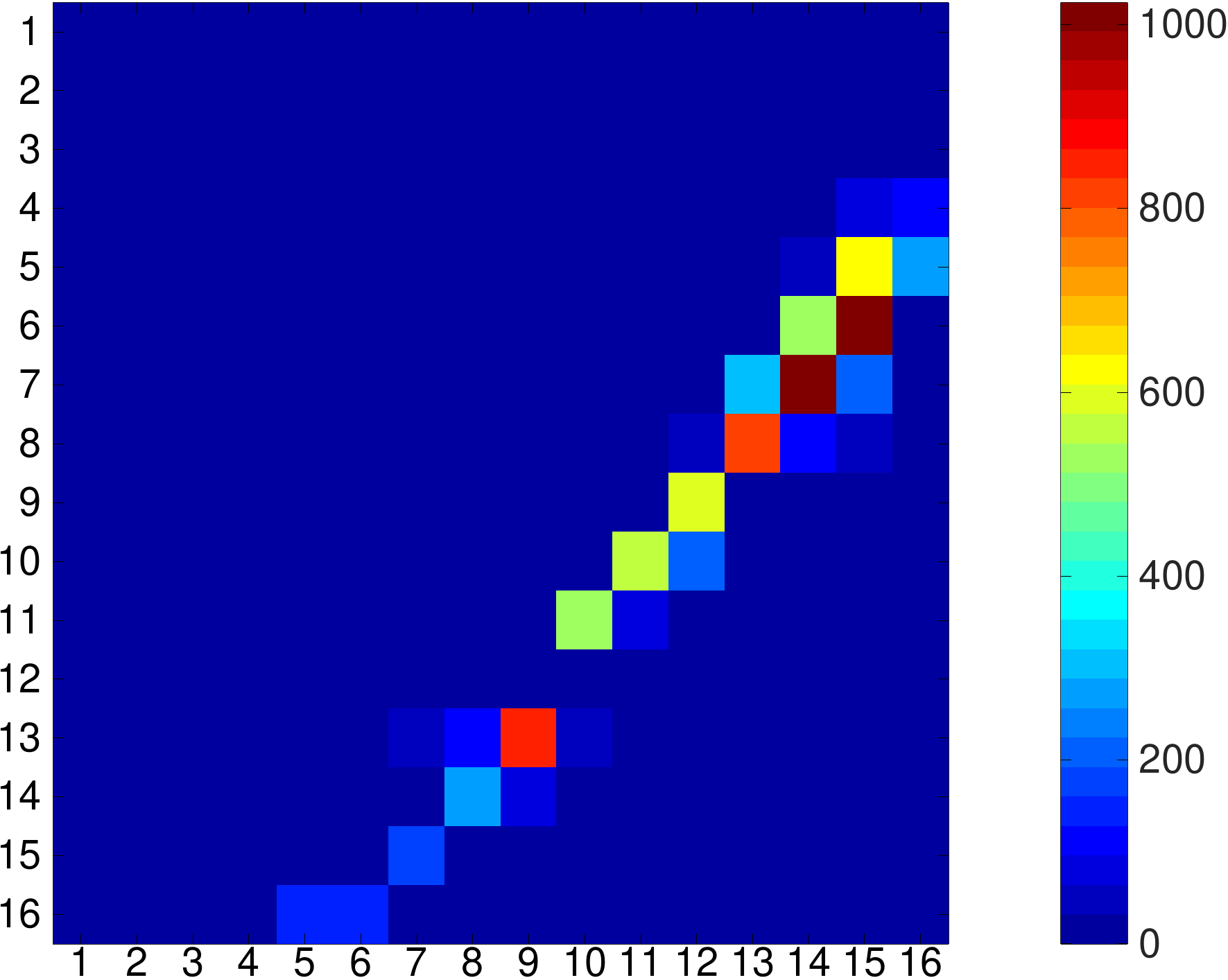}

	\caption{Track-like event registered on September 2, 2016. Top:
	waveforms of nine PMTs that demonstrated the highest ADC counts.
	Bottom: snapshot of the focal plane at the moment of maximum ADC
	counts}
\end{figure}
	
Another impressive example of the measurements is registration
of events with complicated spatial and temporal dynamics. An example
is shown in Fig.~15. The event was registered on December~12, 2016,
near Australia. An arc-like shape of the track made by the brightest
PMTs and the speed of development support the hypothesis that this
was an elve, which represent the most common type of TLEs.

%15 
\begin{figure}[!ht]
	\cfig{1}{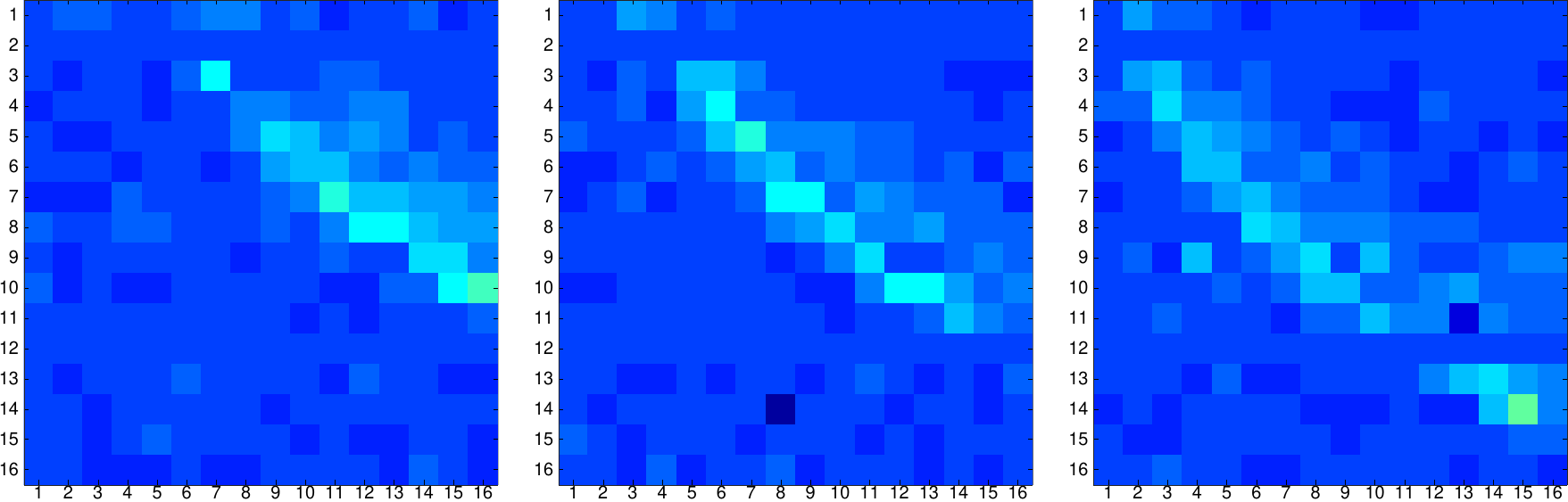}

	\caption{Three snapshots of the focal surface of an elve-like event
	registered on December~12, 2016, near Australia. The ADC counts are
	scaled with respect to individual PMT gains. The time step between
	the snapshots is around 20~$\mu$s}

\end{figure}

\section{Conclusion}

The TUS orbital detector is the first EECR fluorescence telescope
looking down on the earth from on board a satellite. Its aim is to
determine the signal-to-noise ratio for such apparatus, and the results
of its measurements will be used in preparing large-scale space EECR
detectors such as the KLYPVE and JEM-EUSO. 

The fluorescence TUS detector will also give unique data on low
intensity transient atmosphere events and transient luminous events
caused by electric charges, meteoroids, and dust grains.

\section*{Acknowledgements}

The work was partially supported by ROSCOSMOS grants and by RFFI grants
No.\ 16-29-13065 and No.\ 15-35-21038. I.H.~Park was supported by the
National Research Foundation grant funded by MSIP of Korea (No.\
2015R1A2A1A01006870).

\section*{References}
\begin{description}

	\item Abraham, J., Abreu P., Aglietta M. et al. (Pierre Auger
		Collaboration), Trigger and Aperture of the Surface Detector Array
		of the Pierre Auger Observatory. Nucl. Instrum. Methods A
		{\bf613}, 29--39 (2010).

	\item Abrashkin, V., V. Alexandrov, Y. Arakcheev et al., Space
		detector TUS for extreme energy cosmic ray study. Nuclear Physics
		B - Proceedings Supplements, {\bf166}(0), 68-71 (2007).

	\item Aleksandrov V.V., D.I. Bugrov, G.K. Garipov et al., A project
		of investigating the most energetic cosmic rays on the Russian
		segment of the International Space Station. Moscow University
		Physics Bulletin.  {\bf55} (6), 44 (2000).

	\item Benson, R. and Linsley, J., Satellite observation of cosmic ray
		air showers, In International Cosmic Ray Conference, 17th, Paris,
		France, July 13--25, Conference Papers, {\bf8} (1981).

	\item Fenu F., T. Mernik, A. Santangelo et al. ICRC-32 (Beijing)
		(2011).

	\item Garipov G.K., L.A. Gorshkov, B.A. Khrenov et al, AIP Conference
		Proceedings. {\bf433}, 403--417 (1998).

	\item Garipov G.K., B.A. Khrenov, P.A. Klimov et al., Global
		transients in ultraviolet and red-infrared ranges from data of
		Universitetsky-Tatiana-2 satellite. J. Geophys. Res., {\bf118}
		(2), 370--379 (2013).

	\item Grinyuk A.A., A.V. Tkachenko, L.G. Tkachev, TUS Collaboration.
		The TUS orbital detector optical system and trigger simulation.
		Journal of Physics: Conference Series, {\bf409}, Issue 1, article
		id. 012105 (2013).

	\item Gurevich A.V., K.P. Zybin, Runaway breakdown and electric
		discharges in thunderstorms. PHYS-USP, {\bf44} (11), 1119--1140
		(2001).

	\item Khrenov B.A., M.I. Panasyuk, V.V. Alexandrov et al., Space
		Program KOSMOTEPETL (Projects KLYPVE and TUS) for the Study of
		Extremely High Energy Cosmic Rays. AIP Conf. Proc. {\bf566}, 57
		(2001). 

	\item Khrenov B.A., V.P. Stulov, Detection of meteors and
		sub-relativistic dust grains by the fluorescence detectors of
		ultra high energy cosmic rays. Advances in Space Research, {\bf37}
		(10), 1868--1875 (2006).

	\item Krizmanic J.F., J.W. Mitchell, R.S. Streitmatter for the OWL
		Collaboration, Optimization of the Orbiting Wide-angle Light
		Collectors (OWL) Mission for Charged-Particle and Neutrino
		Astronomy, Proc. 33rd ICRC, Rio de Janeiro, Brazil, paper No. 1085
		(2013).

	\item Klimov P.A. PhD Thesis, SINP MSU (2009) (in Russian).

	\item Klimov P.A., Zotov M.Yu., Chirskaya N.P. et al., Preliminary
		results from the TUS ultra-high energy cosmic ray orbital
		telescope: Registration of low-energy particles passing through
		the photodetector.  Bulletin of the Russian Academy of Sciences:
		Physics, {\bf81} (4), 407--409, (2017).

	\item Morozenko V.S. PhD Thesis, SINP MSU (2014) (in Russian).

	\item Panasyuk M.I., M. Casolino M., G.K. Garipov et al., The current
		status of orbital experiments for UHECR studies, J. Phys. Conf.
		Series, {\bf632} (1), 012097 (2015).

	\item Panasyuk M.I., S.I. Svertilov, V.V. Bogomolov et al., RELEC
		mission: Relativistic electron precipitation and TLE study
		on-board small spacecraft. Advances in Space Research, {\bf57}
		(3), 835--849 (2016).

	\item Pasko V.P., Y. Yoav, K. Cheng-Ling, Lightning Related Transient
		Luminous Events at High Altitude in the Earth's Atmosphere:
		Phenomenology, Mechanisms and Effects. Space Sci. Rev. {\bf168}
		(1), 475--516 (2012).

	\item Sadovnichy V.A., M.I. Panasyuk, I.V. Yashin et. al.,
		Investigations of the space environment aboard the
		Universitetsky-Tat'yana and Universitetsky-Tat'yana-2
		microsatellites. Solar System Research, {\bf45} (1), 3--29 (2011).

	\item Scarsi L. Extreme Universe Space Observatory (EUSO). Proc.
		First Airwatch Symposium, Catania; AIP CP, {\bf433}, 42 (1997).

	\item Stecker F.W., J.F. Krizmanic, L.M. Barbier et al., Observing
		the Ultrahigh Energy Universe with OWL Eyes, Nucl. Phys. B (Proc.
		Supp.), {\bf136}, 433--438 (2004).

	\item Takahashi Y. and The JEM-EUSO Collaboration, The JEM-EUSO
		mission, New Journal of Physics, {\bf11}, 065009 (2009).

	\item Vedenkin N.N., G.K. Garipov, P.A. Klimov et al., Atmospheric
		ultraviolet and red-infrared flashes from Universitetsky-Tatiana-2
		satellite data. Journal of Experimental and Theoretical Physics,
		{\bf113} (5), 781--790 (2011).

	\item Voss H.D., M. Walt, W. L. Imhof, J. Mobilia, U.S. Inan,
		Satellite observations of lightning-induced electron precipitation.  J.
		Geophys.  Res. {\bf103} (A6.), 11725 (1998).  \end{description}

\end{document}